\documentclass[final]{svjour3}
\usepackage[dvipdfmx]{graphicx}
\usepackage{rotating}
\usepackage{amssymb}
\usepackage{mathptmx}
\makeatletter
\begin{document}

\newcommand{\hdblarrow}{H\makebox[0.9ex][l]{$\downdownarrows$}-}
\title{Neutrino mass measurements using cryogenic detectors}

\author{L. Gastaldo}

\institute{Kirchhoff Institute for Physics, Heidelberg University,\\ Heidelberg, Germany\\ Tel. +49 6221 549886:\\
\email{Loredana.Gastaldo@kip.uni-heidelberg.de}}

\maketitle

\begin{abstract}

The determination of the absolute mass scale of neutrinos is one of the most important challenges in Particle Physics. The shape of the endpoint region of $\beta$-decay and electron capture (EC) spectra depends on the phase space factor, which, in turn, is function of the neutrino mass eigenstates. High energy resolution and high statistics measurements of $\beta$- and EC spectra are therefore considered a model-independent way for the determination of the neutrino mass scale. Since almost four decades, low temperature microcalorimeters are used for the measurement of low energy $\beta$- and EC spectra. The first efforts were focused on the development of large arrays for the measurement of the $^{187}$Re $\beta$-spectrum. In the last ten years, the attention moved to EC of $^{163}$Ho. This choice was mainly motivated by the very good performance which could be achieved with low temperature microcalorimeters enclosing $^{163}$Ho with respect to microcalorimeters with absorber containing $^{187}$Re. The development of low temperature microcalorimeters for the measurement of the finite neutrino mass is discussed and, in particular, the reasons for moving from $^{187}$Re to $^{163}$Ho. The possibility to reach sub-eV sensitivity on the effective electron neutrino mass with $^{163}$Ho, thanks to the multiplexing of large microcalorimeter arrays is demonstrated. In conclusion, an overview on other nuclides which have been proposed as good candidates, motivated by the excellent performance of low temperature microcalorimeters is presented.


\end{abstract}

\section{Introduction}

The absolute neutrino mass scale is one of the most important questions in physics and possibly it is the key to move beyond the standard model of particle physics. Three approaches are considered to determine the neutrino mass scale: 
\begin{itemize}
	 \item the study of the distribution of matter in the Universe which provides information on the sum of neutrino mass states. The most stringent limit $\sum_{\mathrm{i}}m_{\mathrm{i}}\,<\,120$ meV, where $m_{\mathrm{i}}$ are the values of the single mass eigenstates, has been published by the Planck collaboration \cite{Planck_2018}. This value was obtained by combining studies on the cosmic microwave background anisotropies and the distribution of large structure in the Universe.
	 \item the existence of neutrinoless double beta decay, a lepton number violating process in which two neutrons turn into two protons with the emission of two electrons, can be related to the effective neutrino Majorana mass assuming that a light Majorana neutrino is exchanged between the two nucleons. In this case, the halflife of the process is proportional to square of the effective Majorana mass. Several experiments have been and continue to be developed with the aim to discover such a decay. Presently halflife for neutrinoless double beta decay of the order of $10^{26}$ years have been tested leading to very small upper limits on the effective Majorana mass $m_{\beta\beta}= \left| \sum_i U_{\mathrm{ei}}^2 m_{\mathrm{i}} \right|<79 - 180$ meV at 90$\%$ \cite{Agostini_2020,betabeta}, where $U_{\mathrm{ei}}$ are the elements of the neutrino mixing matrix (Pontecorvo-Maki-Nakagawa-Sakata PMNS matrix)
	 \item The shape of the endpoint region of low energy electron capture and beta decay spectra depends on the absolute mass of neutrino mass states \cite{kinematic}. Recently, the KATRIN collaboration, measuring the $^3$H $\beta$-spectrum, has released a new upper limit on the effective electron anti-neutrino mass $m_{\nu}<0.8$ eV 95$\%$ C.L. \cite{Aker_2021} and the ECHo collaboration has published a limit for the effective mass of the electron neutrino $m_{\nu}<150$ eV at 90$\%$ C.L. \cite{Velte_2019}.
\end{itemize} 
\noindent Low temperature detectors have played and will play a very important role in experiments related to all the three approaches to determine the absolute neutrino mass scale. As examples, the spiderweb detectors mounted on the Planck satellite \cite{spiderweb} and the large calorimeters developed for the search of neutrinoless double beta decay  \cite{CUORE,AMoRE,CUPID}. In the following the use of low temperature micro-calorimeters for the high energy resolution and high statistics measurement of beta and electron capture spectra will be discussed together with the achievable sensitivity on the effective neutrino mass of running and planned experiments.   

\section{Calorimetric measurements of low energy $\beta$- and EC-spectra}
A beta decay is a process in which an electron and an electron antineutrino are emitted while a neutron in a nucleus turns into a proton. Electron capture is a process in which an electron from the inner shells is captured by the nucleus and an electron neutrino is emitted while a proton turns into a neutron. The total energy available to these decays is a fixed parameter $Q$, defined by the difference between the mass of the parent and daughter atom. This energy is shared between the emitted (anti-)neutrino and the emitted electron in beta decays or atomic excitations in electron captures (in both decays also a small fraction of energy is going in the nuclear recoil). The emitted electron (anti-)neutrinos will escape detection and, therefore, both the spectral shape of beta and electron capture decay contain the phase space term which carries the dependency of the finite neutrino masses:
\begin{equation}
	\frac{dN}{dE} \propto \sum_i \left| U_{\mathrm{ei}}\right|^2 (Q-E)^2\sqrt{1-\frac{m_i^2}{(Q-E)^2}} \rightarrowtail \frac{dN}{dE} \propto (Q-E)^2\sqrt{1-\frac{m(\nu)}{(Q-E)^2}}
\end{equation}
The right expression derives from the left one considering that available detectors have still not reached energy resolution of the order of the difference between the neutrino mass eigenstates. Therefore the quantity that can be determined by on-going and next future experiments is the effective neutrino mass $m(\nu)$, defined as \cite{Huang_2020}:
\begin{equation}
	m(\nu)^2 = \sum_i  \left| U_{\mathrm{ei}}\right|^2 m_i^2
\end{equation}
Therefore, the analysis of the endpoint region of high energy resolution and high statistic beta and electron capture spectra is an ideal approach for the determination of the neutrino mass scale only using energy and momentum conservation. This is the idea Fermi has proposed just after having developed the theory of beta decay \cite{Fermi}. \\
Nowadays two nuclides are used in experiments for the determination of the effective electron neutrino mass $^3$H and $^{163}$Ho \cite{kinematic}. The KATRIN experiments, measuring the high energy part of the $^3$H $\beta$-spectrum using a huge electron spectrometer, has recently released a new upper limit on the effective electron anti-neutrino mass $m_{\nu}<0.8$ eV 95$\%$ C.L. \cite{Aker_2021} and the collaboration is moving forwards to reach a sensitivity as low as 0.2 eV. Project 8 \cite{Esfahani_2017} and PTOLEMY \cite{Betti_2019} are two experiments which plan to exceed in the future the sensitivity foreseen for KATRIN with innovative technologies. The ECHo \cite{Gastaldo_EPJST} and HOLMES \cite{Alpert_2015} collaborations are planning to reach similar sensitivity as the KATRIN experiment by the analysis of high energy resolution $^{163}$Ho spectra measured using low temperature microcalorimeters enclosing the radioactive source.\\
Experiments using low temperature detectors in which the radioactive source is surrounded by enough material so that only the neutrino can escape the detector volume present some advantages with respect to experiments in which the source is positioned outside the detector. In fact, the 100$\%$ quantum efficiency for all the energy released in the weak nuclear decay besides the one taken away by the neutrinos and including also the fraction of energy in nuclear recoil, implies that no corrections due to energy loses either in the source itself or in non-active parts of the detector have to be considered. In addition, due to the fast thermalization processes in metallic particle absorbers \cite{Kozorezov}, no corrections due to possible energy stored in excited states have to be applied. Nevertheless, calorimetric measurements also face intrisic problems which are related to the fact that, being the source part of the detector, each single decay leads to a signal. Therefore, there is no possibility to filter out only events close to the endpoint region as for the KATRIN electron spectrometer. In addition, the measurement will be affected by an intrinsic background given by unresolved pile-up events. The fraction of events leading to unresolved pile-up is given by the product of the time resolution of the detector and the source activity. Therefore, reducing the pile-up background would mean reducing the enclosed activity, once the time resolution is fixed. Another important aspect which should be considered when designing low temperature microcalorimeters containing a radioactive source is the contribution that the source atoms will give to the total detector heat capacity. Infact, the intrinsic energy resolution is scaling with the square root of the detector heat capacity and therefore a careful design optimization needs to be performed in which both source activity and quantum efficiency are maximized with the aim to keep the energy resolution below a given threshold. \\
While nowadays only low temperature detectors enclosing $^{163}$Ho are used in experiments for the determination of the neutrino mass scale, the first experiments were actually dedicated to the study of $^{187}$Re beta spectrum \cite{Vitale}. Presently, the availability of new high precision measurements of $Q$-values for nuclides undergoing electron capture and beta decays, has driven the attention to the possibility of studying decays to excited nuclear states with low temperature detectors.   
\section{$^{187}$Re based experiments}
The isotope $^{187}$Re represents about the 63$\%$ of natural rhenium. Besides the very long halflife of about $4\times 10^{10}$ years, implying that for an activity of 1 Bq about 1 mg of natural renium is required, the very low energy available for the beta decay $Q = 2.479(13)$ eV \cite{Filianin_2021}, the lowest among all beta decays to ground nuclear states, makes $^{187}$Re one of the most interesting candidates to be used in experiments for the determination of the neutrino mass scale. In fact, such a low $Q$-values implies that in the last 10 eV below the end-point (assuming vanishing neutrino mass) a fraction of events of about $10^{-7}$ is expected.
Very precise spectra have been measured by two collaborations MANU \cite{MANU} and MIBETA \cite{MIBETA} which provided promising upper limits for the effective electron anti-neutrino mass: $m(\nu)<26$ eV at 95$\%$ C.L. and $m(\nu)<15$ eV at 90$\%$ C.L. respectively. The spectra acquired in the two experiments were measured with quite different calorimeter concepts. In MANU one detector consisting of a 1.6 mg superconducting rhenium absorber was used in order to maximize the amount of $^{187}$Re while having a low heat capacity. The rhenium crystal was coupled to a NTD-Ge sensor to monitor the temperature. In MIBETA instead the measurement was performed with ten pixels consisting of silver perrhenate crystals (AgReO$_4$) of about 0.3 mg coupled to silicon thermistors. 
The analysis of the spectra acquired in both experiments allowed also to precise characterize the influence on the spectral shape due to the coherent interference of the electron emitted in the $^{187}$Re beta decay with the lattice. The so called Beta Environmental Fine Structure \cite{Koonin} is an oscillating pattern superimposed to the beta spectrum which potentially could induce systematic uncertainties in the analysis of the endpoint region of the spectrum.  \\
In 2006 the MANU and MIBETA collaborations together with other experts in the development of large arrays of low temperature detectors formed a new collaboration with name MARE with the goal to develop a large experiment    
able to reach sub-eV sensitivity on the effective electron neutrino mass using $^{187}$Re \cite{MARE}. Nevertheless, after half a decade of detector optmization in which superconducting rhenium was coupled to different sensors as Transition Edge Sensors (TESs) and Metallic Magnetic Calorimeters (MMCs) and specific tests on AgReO$_4$ to model the asymmetric detector response were performed, neither the envisioned energy resolution of the order of few eV FWHM could not be achieved nor a precise description of the detector response for Re-based absorbers \cite{Ranitzsch,Sisti}. Presently no groups are anymore working on detector optimization for a $^{187}$Re experiment.
\section{$^{163}$Ho based experiments}
The idea to use the nuclide $^{163}$Ho for the determination of the effective neutrino mass has been proposed in the early eighties \cite{ADRML}. The halflife of $^{163}$Ho has been measured to be 4570 years \cite{Baisden} and the energy available to the decay has been determined to be about 2.83 keV \cite{Velte_2019,Eliseev}, which is the lowest for all nuclides undergoing electron capture the ground nuclear state. The use of low temperature microcalorimeters enclosing the radioactive source for the measurement of the $^{163}$Ho spectrum was demonstrated in 1996 \cite{Meunier}. The moderate performance of the detector, the difficulty to produce high purity $^{163}$Ho sources together with the promising results obtained with $^{187}$Re lead to a decrease of interest in $^{163}$Ho-based experiments.\\
A renewal of the interest to determine the effective electron neutrino mass with a large scale $^{163}$Ho experiment started after the publication of the high energy resolution $^{163}$Ho spectrum measured using low temperature MMCs with $^{163}$Ho enclosed in a gold absorber \cite{Gastaldo_NIMA,Ranitzsch_PRL}. These very good results represented the start of the ECHo experiment \cite{Gastaldo_EPJST} which is conceived with the goal to achieve sub-eV sensitivity on the effective electron neutrino mass using large arrays of multiplexed MMCs hosting $^{163}$Ho. A second large collaboration is also pursuing the same aim, the HOLMES collaboration \cite{Alpert_2015} which plan to use large arrays of multiplexed Mo/Cu TES on SiN$_x$ membranes \cite{Giachero_2021}.\\ 
The requirements to be considered for designing such an experiment can be motivated by analysing the expected spectral shape. The fact that the fraction of events in the last 10 eV below the $Q$-value (assuming vanishing neutrino masses) is just of the order of $10^{-9}$ implies that a total number of $^{163}$Ho events of the order of $10^{14}$ need to be acquired, which, in turn, implies that a $^{163}$Ho source with activity of the order of MBq needs to be enclosed in microcalorimeters characterized by energy resolution below 5 eV FWHM. The maximum amount of $^{163}$Ho which can be hosted in a single detector is then limited by the allowed unresolved pile-up fraction and the contribution of the holmium atoms to the detector heat capacity. For the ECHo experiment the optimal activity has been determined to be 10 Bq per pixel which, combined with a time resolution of MMC well-below 1 $\mu$s would lead to an unresolved pile-up fraction well-below $10^{-5}$ and, at the same time lead to an additional heat capacity of less than 50$\%$ of the one of gold absorber and Ag:Er sensor when the detectors are operated at a temperature of 20 mK. In \cite{Herbst} the specific heat per Ho atom in gold and silver has been measured as a function of Ho concentration and it was shown that for temperature above 100 mK the temperature dependence of the Ho specific heat follows that one measured for pure Ho samples, showing a Schottky anomaly at about 250 mK. This fact might induce some issues for the optimization of TES-based detector for the HOLMES experiment for which an activity of 300 Bq per pixel and an operational temperature around 80 mK are presently considered.

\noindent Both collaborations have decided to produce $^{163}$Ho at nuclear reactor using enriched $^{162}$Er targets. Neutron irradiation has been performed for both experiment at the high flux nuclear reactor at ILL in Grenoble \cite{ILL}. In early days of R\&D for $^{163}$Ho source production also other methods have been considered as discussed in \cite{Engle} and a sample of $^{163}$Ho produced via proton irradiation of a Dy target was used in a proof of principle experiment by the NuMECS collaboration \cite{Croce}. The decision to use neutron activation of $^{162}$Er enriched samples has been motivated by the high production rate and the availability of a very efficient chemical purification, which removes all elements besides holmium \cite{Dorrer,Heinz}. Both the ECHo and HOLMES collaboration have produced and chemically separated enough $^{163}$Ho to be used in the planned phases of the experiments. \\
The enclosing of $^{163}$Ho in absorber of microcalorimeters is obtained with ion-implantation of a mass separated 163 u beam generated from a target prepared with a portion of the chemically purified holmium fraction. The ion-implantation has several advantages. In combination with a bending magnet it provides an efficient selectivity for the mass 163 u removing both contamination traces of target material or other elements. Of particular importance is the removal of the holmium isotope $^{166m}$Ho which undergoes beta decay with a halflife of about 1200 years and is present in the chemically separated holmium fraction at the level of $10^{-4}$ with respect to $^{163}$Ho \cite{Dorrer} The presence of $^{166m}$Ho in the detectors would generate a non-negligible background in the region of interest of the $^{163}$Ho spectrum. With an acceleration potential around 30 kV, the implantation volume in the absorber is well-defined allowing for a safe dimensioning of the first absorber layer, where the ions are implanted, and second absorber layer, which covers the implantation area. The ECHo collaboration has demonstrated that detectors having implanted $^{163}$Ho in the absorbers have not degraded performance and that the fabrication of the absorbers is reliable \cite{Gastaldo_NIMA,Mantegazzini_2022}. While the first $^{163}$Ho implantation processes for the ECHo experiment have been performed at ISOLDE-CERN \cite{ISOLDE}, more recent implantation processes have been performed at the RISIKO facility at Mainz University which has been optimized to perfectly suite the requirements for $^{163}$Ho implantation in arrays of MMCs \cite{Kieck1,Kieck2}. The HOLMES collaboration has decided to build a mass separation - ion implanter dedicated for the experiment. The system is located at Genoa University and is presently under commissioning \cite{Gallucci}.  \\
The ECHo collaboration has optimized the MMC design with $^{163}$Ho loaded absorbers over the last ten years, demonstrating that energy resolution of a few eV FWHM can be obtained with detectors having about 1 Bq of $^{163}$Ho activity \cite{Gastaldo_NIMA,Mantegazzini_2022,Velte_recent_results,Griedel_2022} and has been developing a microwave multiplexing design optimized for MMC features \cite{Kempf,Wegner}. After the first proof of concept phase which brought to the publication of a new limit for the effective electron neutrino mass of 150 eV \cite{Velte_2019}, improving after about 30 year the limit obtained by Springer et al. with the measurement of the x-ray spectrum of $^{163}$Ho \cite{Springer_1987}, the ECHo collaboration has completed in 2020 the data acquisition for the ECHo-1k phase of the experiment collecting about $10^8$ $^{163}$Ho events. Data analysis is presently on-going. At the same time the preparation for the ECHo-100k phase of the experiment has started. For this phase about 12000 MMC channels with 10 Bq $^{163}$Ho per pixel will be readout using SQUID microwave multiplexing \cite{Kempf,Wegner}. The fabrication of the newly designed ECHo-100k wafer is proceeding \cite{Griedel_2022} as well as R\&D for wafer scale implantation.\\
The HOLMES collaboration has demonstrated very good performance for multiplexed TES arrays designed for the experiment, but still has to demonstrate that the same performance are kept also for TES hosting $^{163}$Ho \cite{Giachero_2021,HOLMES_Det}. The HOLMES experiment will consist of 1000 TES detector each loaded with about 300 Bq $^{163}$Ho. This very agressive plan aiming at acquiring a very high number of $^{163}$Ho events in a short time and with a moderate number of channels looks first quite challenging, but could still be demonstrated to be the optimal choice for the experiment.
Both collaborations have also developed dedicated data analysis algorithms for efficiently reduce data acquired with large microcalorimeter arrays \cite{Hammann,HOLMES_DataRed}. In particular the HOLMES collaboration has demonstrated an efficient pile-up discrimination which is of utmost importance for detector designed to have about 300 Bq of $^{163}$Ho activity \cite{pu_reduction}. First investigations to determine the background model in the ECHo experiment have been also performed showing that the goal of having the background level due to natural radiactivity and muon related events can be kept below the unresolved pile-up level \cite{Goeggelmann1,Goeggelmann2}. In the last years also new calculation to determine the theoretical shape of the $^{163}$Ho spectrum have been performed, bringing the agreement between theory and experimental data to a much better level \cite{Brass1,Brass2}.   

\section{$\beta$-decay and electron capture to excited nuclear states}
For the nuclides discussed so far, decays to the ground nuclear states have been considered. A possibility for studying electron capture and beta decays with $Q$-values lower than the one of $^{163}$Ho and $^{187}$Re is to consider decays to excited nuclear states with excitation energy close to the $Q$-value for the decay to the ground state. If the lifetime of the excited nuclear state is of the order of $\mu$s or shorter, it would be possible to recognize decays to the excited nuclear state by measuring in coincidence the emitted gamma and the electron energy in case of beta decay or atomic de-excitation energy in case of electron capture. This possibility would then open the way to study beta and electron capture spectra where a large fraction of decay is in the region of interest below the endpoint. \\
The requirements to perform such an experiments, once a proper candidate has been identified, would be: a low temperature micro-calorimeter, enclosing the nuclide of interest as the one developed for $^{163}$Ho-based experiments and large crystal calorimeters, similar to the one developed for dark matter searches or neutrinoless double beta decay searches surrounding the micro-calorimeter array. The small micro-calorimeters have the task to measure the energy of the beta electron or atomic de-excitation (and nuclear recoil) following beta decay and electron capture respectively, while the large crystal detectors have the task to detect the emitted gamma to identify the decay to the excited nuclear state based on coincidence criterium.   
Therefore, from the technology point of view, suitable detectors are available, the next step is to understand if there are nuclides which could be used for competitive experiments. \\
The first beta decay to an excited nuclear state which was generating interest for a possible use in experiments for the direct determination of the effective electron neutrino mass is the one of $^{115}$In \cite{Cattadori,Mount}. The indium isotope $^{115}$In has a 96\% natural abundance, therefore no problem for isotope production/enrichment, a $Q$-value to the nuclear ground state of 497.489(10) keV and an extremely long halflife of $4.4 \times 10^{14}$ years. The nuclear excited state of interest is at 497.334(22) keV, giving then an energy available to the beta decay of only 155 eV.
The branching ratio for decays to the excited state is about $1.2 \times 10^{-4}$ and the excited state halflife is of only 11 ps, ensuring the possibility to have precise coincidence measurements. The challenge of a $^{115}$In-based experiment is the amount of indium which needs to be used for absorber of microcalorimeters featuring eV-scale energy resolution. For each Bq about $2\times 10^{22}$ $^{115}$In atoms are required which correponds to about 30 g of indium, therefore for having 1 decay per second to the excited nuclear state about $2\times 10^{26}$ $^{115}$In atoms or 300 kg indium need to be used. This size of active material is similar to the one used in the search for neutrinoless double beta decay.\\
In 2020 a new high precision Penning Trap Mass Spectrometry measurement was pointing out the case of $^{135}$Cs \cite{deRoubin}. This cesium isotope undergoes a beta decay to $^{135}$Ba with a halflife of about $1.3\times 10^{6}$ years, about $10^9$ times shorter than the one of $^{115}$In, and need to be artificially produced. The energy available to the decay is 268.66(30) keV and the energy of the excited nuclear state of interest, with a branching ratio of $1.6 \times 10^{-6}$ is 268.218(20) keV giving a $Q$-value for the beta decay to the excited nuclear state of only 440 eV. The reason why $^{135}$Cs cannot be used in an experiment for the measurement of the beta spectrum of decays to the nuclear excited state is that the halflife of the excited nuclear level is 28.11 h, more than one day! This long decay time together with the branching ratio would make the identification of enough decays to the excited state more than challenging.\\
A last nuclides which is worthwhile to discuss is the artifically produced $^{159}$Dy \cite{Ge}. $^{159}$Dy undergoes an electron capture with a $Q$-value of 364.73(19) keV and a quite short halflife of only 144 days. With a branching ratio of about $7 \times 10^{-4}$, $^{159}$Dy decays to an excited nuclear state at 363.5449(14) keV giving a $Q$-value for the electron capture to that excited nuclear state of 1.19 keV. Among the three nuclides here discussed $^{159}$Dy represents the most meaningful candidate to be used in an experiment for the determination of the effective electron neutrino mass. In \cite{Ge} a comparison of the decay rate in the endpoint region of the electron capture spectrum of $^{159}$Dy and $^{163}$Ho was discussed. For normalized spectra the rate for $^{159}$Dy is slightly larger than the one for $^{163}$Ho. On the other hand, in order to acquire a given number of $^{159}$Dy decays to the excited nuclear state, more that $10^3$ times more decays would be detected in the microcalorimeters. This implies that if the rate of decays in the region of interest is normalized by the total number of decays occurring in the calorimeters, the best scenario is still achieved using $^{163}$Ho.\\
In summary, searches for low $Q$-value electron capture and beta decay to excited nuclear states is very important and might lead to the discovery of electron capture candidates having an atomic excitation energy very close to the maximum energy available to the decay, which would strongly enhance the rate at the end point region.   

\section{Conclusions}
\noindent Even if the kinematic approach, consisting in the analysis of beta and electron capture spectra, provides up to now the worst limit for the absolute neutrino mass scale, it has the advantage to be model independent since it relies only on energy and momentum conservation. It is clear that, in case of any claim of discovery of a finite value either for the sum of neutrino masses from cosmology or for the effective Majorana mass from the observation of neutrinoless double beta decay, a confirmation from a direct kinematic measurement would be necessary to confirm the discovery and to better constrain the neutrino mass scale. \\
The optimization of low temperature detectors with absorber enclosing the nuclide of interest made possible the development of experiments with the potential to achieve sub-eV sensitivity on the effective electron neutrino mass and a realistic perspective to reach in the near future similar sensitivities of $^3$H-based experiments. The ECHo experiment using $^{163}$Ho has already provided an improved limit on the effective electron neutrino mass and the R\&D performed by the ECHo and HOLMES collaborations demonstrated the feasibility of large scale $^{163}$Ho-based experiments. In parallel, high precision Penning Trap Mass Spectrometry measurements are performed to reduce uncertainties on the $Q$-value for nuclides undergoing electron capture and beta decays to identify if excited nuclear levels, which are already known via gamma spectroscopy, would be at about 1 keV or below from the $Q$-value. Those nuclide might be avantageous with respect to decays to the ground state since a higher count rate at the endpoint region could be obtained, reducing cost and running time of the experiment. Presently $^{163}$Ho can still be considered the best nuclides to be used in direct neutrino mass determination using low temperature detectors. 

\begin{acknowledgements}
LG would like to thank members of the MARE, ECHo and HOLMES collaboration for inspiring discussions
\end{acknowledgements}

%\pagebreak


\begin{thebibliography}{99}

\bibitem{Planck_2018} 
Planck Collaboration, \emph{A$\&$A} \textbf{641} (2020) A6

\bibitem{Agostini_2020}
M. Agostini et al., (The GERDA Collaboration), \emph{Phys. Rev. Lett.} \textbf{125} (2020) 252502

\bibitem{betabeta}
M. Agostini et al., arXiv:2202.01787 [hep-ex]

\bibitem{kinematic}
L. Gastaldo, \emph{PoS (NOW2016)} (2017) 060, https://pos.sissa.it/283/

\bibitem{Aker_2021}
The KATRIN Collaboration, \emph{Nature Physics} \textbf{18} (2022) 160

\bibitem{Velte_2019}
C. Velte et al., \emph{The European Physical Journal C} \textbf{79} (2019) 1026

\bibitem{spiderweb}
M. Yun et al., \emph{Journal of vacuum science \& technology} \textbf{B 22(1)} (2004) 220

\bibitem{CUORE}
CUORE Collaboration, arXiv:2104.06906 [nucl-ex] 

\bibitem{AMoRE}
V. Alenkov et al., \emph{The European Physical Journal C} \textbf{79} (2019) 791 

\bibitem{CUPID}
The CUPID Interest Group, arXiv:1504.03612 [physics.ins-det]

\bibitem{Huang_2020} 
G.-Y. Huang et al., \emph{Phys. Rev. D} \textbf{101/1} (2020) 016003 

\bibitem{Fermi}
E. Fermi, \emph{Z. Phys.} \textbf{88} (1934) 161

\bibitem{Esfahani_2017}
A. A. Esfahani et al. (Project8 Collaboration), \emph{Journal of Physics G: Nuclear and Particle Physics} \textbf{44/5} (2017) 054004

\bibitem{Betti_2019}
M. G. Betti et al. (PTOLEMY Collaboration), \emph{Journal of Cosmology and Astroparticle Physics} \textbf{2019/07} (2019) 047

\bibitem{Gastaldo_EPJST} 
L. Gastaldo et al. (ECHo Collaboration),\emph{Eur. Phys. J. Spec. Top.} \textbf{226} (2017) 1623

\bibitem{Alpert_2015}
B. Alpert et al., \emph{Eur. Phys. J. C} \textbf{75} (2015) 112

\bibitem{Kozorezov}
A. G. Kozorezov et al., \emph{Journal of Low Temperature Physics} \textbf{167} (2012) 473

\bibitem{Vitale} 
A. Blasi et al., \emph{I.N.F.N./BE-85/2, internal report} (1985)

%\bibitem{Suhonen}
%J. Suhonen et al.,\emph{text} \textbf{text}

\bibitem{Filianin_2021} 
P. Filianin et al.,\emph{Phys. Rev. Lett.} \textbf{127} (2021) 072502

\bibitem{MANU} 
M. Galeazzi et al., \emph{Phys. Review C} \textbf{63/1} (2000) 014302

\bibitem{MIBETA}
M. Sisti et al.,\emph{Nucl. Instrum. Methods Phys. Res. A} \textbf{520} (2004) 125

\bibitem{Koonin} 
S. E. Koonin,\emph{Nature} \textbf{354} (1991) 468

\bibitem{MARE} 
A. Monfardini et al.,\emph{Nucl. Instrum. Methods Phys. Res. A} \textbf{559} (2006) 346

\bibitem{Ranitzsch} 
P. C.-O. Ranitzsch et al.,\emph{J. of Low Temp. Phys} \textbf{167} (2012) 1004
 
\bibitem{Sisti} 
E. Ferri et al.,\emph{The European Physical Journal A} \textbf{48} (2012) 1
 
\bibitem{ADRML} 
A. De Rujula and M. Lusignoli,\emph{Phys. Lett. B} \textbf{118} (1983) 429
 
\bibitem{Baisden} 
P. A. Baisden et al., \emph{Phys. Rev. C (Nuclear Physics)} \textbf{28-1} (1983) 337

\bibitem{Eliseev} 
S. Eliseev et al., \emph{Phys. Rev. Lett.} \textbf{115} (2015) 062501
 
\bibitem{Meunier} 
F. Gatti et al.,\emph{Physics Letters B} \textbf{398} (1997) 415
 
\bibitem{Gastaldo_NIMA} 
L. Gastaldo et al., \emph{Nucl. Inst. Meth. A} \textbf{711} (2013) 150

\bibitem{Ranitzsch_PRL} 
P. C.-O. Ranitzsch et al., \emph{Phys. Rev. Lett.} \textbf{119} (2017) 122501 

\bibitem{Giachero_2021} 
A. Giachero et al.,\emph{IEEE Transactions on Applied Superconductivity} \textbf{31} (2021) 2100205

\bibitem{Herbst} 
M. Herbst et al.,\emph{J. Low Temp. Phys.} \textbf{202} (2021) 106

\bibitem{ILL}
https://www.ill.eu/

\bibitem{Engle} 
J.~W. Engle et al., \emph{Nucl. Instrum. Meth. B} \textbf{311} (2013) 131

\bibitem{Croce} 
M. P. Croce et al., {\it J. Low Temp. Phys.} \textbf{184 - 3} (2016) 938

\bibitem{Dorrer} 
H. Dorrer et al., \emph{Radiochimica Acta} \textbf{106/7} (2018) 535

\bibitem{Heinz} 
S. Heinitz et al., \emph{PLoS ONE} \textbf{13(8)} (2018) e0200910

\bibitem{Mantegazzini_2022} 
F. Mantegazzini et al., accepted for publication in \emph{Nucl. Instrum. Meth. A}, https://doi.org/10.1016/j.nima.2022.166406

\bibitem{ISOLDE} 
E. Kugler, {\it Hyperfine Interact.} \textbf{129} (2000) 23

\bibitem{Kieck1} 
T. Kieck et al., \emph{Review of Scientific Instruments} \textbf{90} (2019) 053304

\bibitem{Kieck2} 
T. Kieck et al., \emph{Nucl.Instrum.Meth.A} \textbf{945} (2019) 162602

\bibitem{Gallucci} 
G. Gallucci et al., \emph{J.Low Temp.Phys.} \textbf{194} (2019) 453

\bibitem{Velte_recent_results} 
C. Hassel et al., \emph{J. Low Temp. Phys.} \textbf{184 - 3} (2016) 910

\bibitem{Griedel_2022} 
M. Griedel et al.,accepted for publication in \emph{J. Low Temp. Phys.} 

\bibitem{Kempf} 
S. Kempf et al.,\emph{AIP Advances} \textbf{7} (2017) 015007

\bibitem{Wegner} 
M. Wegner et al.,\emph{J. Low Temp. Phys.} \textbf{193} (2018) 462

\bibitem{Springer_1987} 
P. T. Springer et al., \emph{Phys. Rev. A} \textbf{35} (1987) 679

\bibitem{HOLMES_Det} 
D.T. Becker et al., \emph{JINST} \textbf{14} (2019) P10035

\bibitem{Hammann} 
R. Hammann et al.,\emph{Eur. Phys. J. C} \textbf{81} (2021) 963

\bibitem{HOLMES_DataRed} 
M. Borghesi et al.,\emph{Eur. Phys. J. C} \textbf{81} (2021) 385

\bibitem{pu_reduction} 
M. Borghesi et al., arXiv:2201.05549 [physics.data-an]

\bibitem{Goeggelmann1} 
A. Goeggelmann et al., \emph{Eur.Phys.J.C} \textbf{81} (2021) 363

\bibitem{Goeggelmann2} 
A. Goeggelmann et al., \emph{Eur.Phys.J.C} \textbf{82} (2022) 139

\bibitem{Brass1} 
M. Brass et al., \emph{Phys. Rev. C} \textbf{97} (2018) 054620 

\bibitem{Brass2} 
M. Brass and M. W. Haverkort, \emph{New Journal of Physics} \textbf{22} (2020) 093018

\bibitem{Cattadori} 
C. M. Cattadori et al.,\emph{Phys. of Atom. Nucl.} \textbf{70} (2007) 127

\bibitem{Mount} 
B. J. Mount et al.,\emph{Phys. Rev. Lett.} \textbf{103} (2009) 122502

\bibitem{deRoubin} 
A. de Roubin et al.,\emph{Phys. Rev. Lett.} \textbf{124} (2020) 222503

\bibitem{Ge} 
Z. Ge et al.,\emph{Phys. Rev. Lett.} \textbf{127} (2021) 272301

\end{thebibliography}
\end{document}